\documentclass[prd,preprint,eqsecnum,amsmath,amssymb,nofootinbib]{revtex4}
\usepackage{graphicx, tikz}
\usepackage{caption, subcaption}
\usepackage{amssymb,amsmath, mathrsfs}
\usepackage{amssymb, graphics, setspace}
\usepackage{hyperref}
\usepackage[all]{xy}
\newcommand{\be}{\begin{equation}}
\newcommand{\ee}{\end{equation}}
\newcommand{\bea}{\begin{eqnarray}}
\newcommand{\eea}{\end{eqnarray}}
\newcommand{\bes}{\begin{subequations}}
\newcommand{\ees}{\end{subequations}}

\begin{document}

\title{The Unruh vacuum and the in-vacuum in the Reissner-Nordstr\"om spacetime
\footnote{This paper is dedicated to Richard Kerner in the occasion of his 80th birthday.}
}
\author{Roberto~Balbinot}
\email{roberto.balbinot@unibo.it}
\affiliation{Dipartimento di Fisica dell'Universit\`a di Bologna and INFN sezione di Bologna, Via Irnerio 46, 40126 Bologna, Italy
%
}
\author{Alessandro~Fabbri}
\email{afabbri@ific.uv.es}
\affiliation{Departamento de F\'isica Te\'orica and IFIC, Universidad de Valencia-CSIC, C. Dr. Moliner 50, 46100 Burjassot, Spain
}

\bigskip\bigskip

\begin{abstract}

The Unruh vacuum is widely used as quantum state to describe black hole evaporation since near the horizon it reproduces the physical state of a quantum field, the so called in-vacuum, in the case the black hole is formed by gravitational collapse.

We examine the relation between these two quantum states in the background spacetime of a Reissner-Nordstr\"om black hole (both extremal and not) highlighting similarities and striking differences. 

\end{abstract}
\maketitle

\section{Introduction}

In Quantum Field Theory in black hole (BH) spacetimes particular relevance is given to the so called `Unruh vacuum'  \cite{Unruh:1976db}. This quantum state is defined on the maximal analytic extension of the BH spacetime by expanding the quantum field operators in ingoing modes modes that are positive frequency with respect to the asymptotic Minkowski time, while the outgoing modes, emerging from the past horizon, are chosen to be positive frequency  with respect to Kruskal time.
This state is expected to describe, at late retarded time, the quantum state of a field in the spacetime of a collapsing body forming a BH. 

Let us make an explicit example in the simple setting of a two-dimensional Schwarzschild spacetime whose metric reads
\be \label{unouno} ds^2=-(1-\frac{2m}{r})dt^2+(1-\frac{2m}{r})^{-1}dr^2=-(1-\frac{2m}{r})dudv\ , \ee
where $m$ is the BH mass and
\bea
&& u=t-r^*,\  \label{unoduea} \\ && v=t+r^* \label{unodueb} \eea
are, respectively, the retarded and advanced Eddington-Finkelstein coordinates. $r^*$ is Regge-Wheeler `tortoise' coordinate
\be \label{unotre}
r^*=\int \frac{dr}{(1-\frac{2m}{r})}=r+2m\ln|\frac{r}{2m}-1| \ . \ee
Kruskal's null coordinates are defined as \cite{kruskal}
\bea \label{unoquattroa} && U=\pm e^{-\kappa u}\ , \\ \label{unoquattrob} && V=\pm \frac{1}{\kappa}e^{\kappa v}\ , \eea
where $\kappa=1/4m$ is the surface gravity of the BH horizon located at $2m$.

\begin{figure}[h]
\centering \includegraphics[angle=-1.5, height=3.5in] {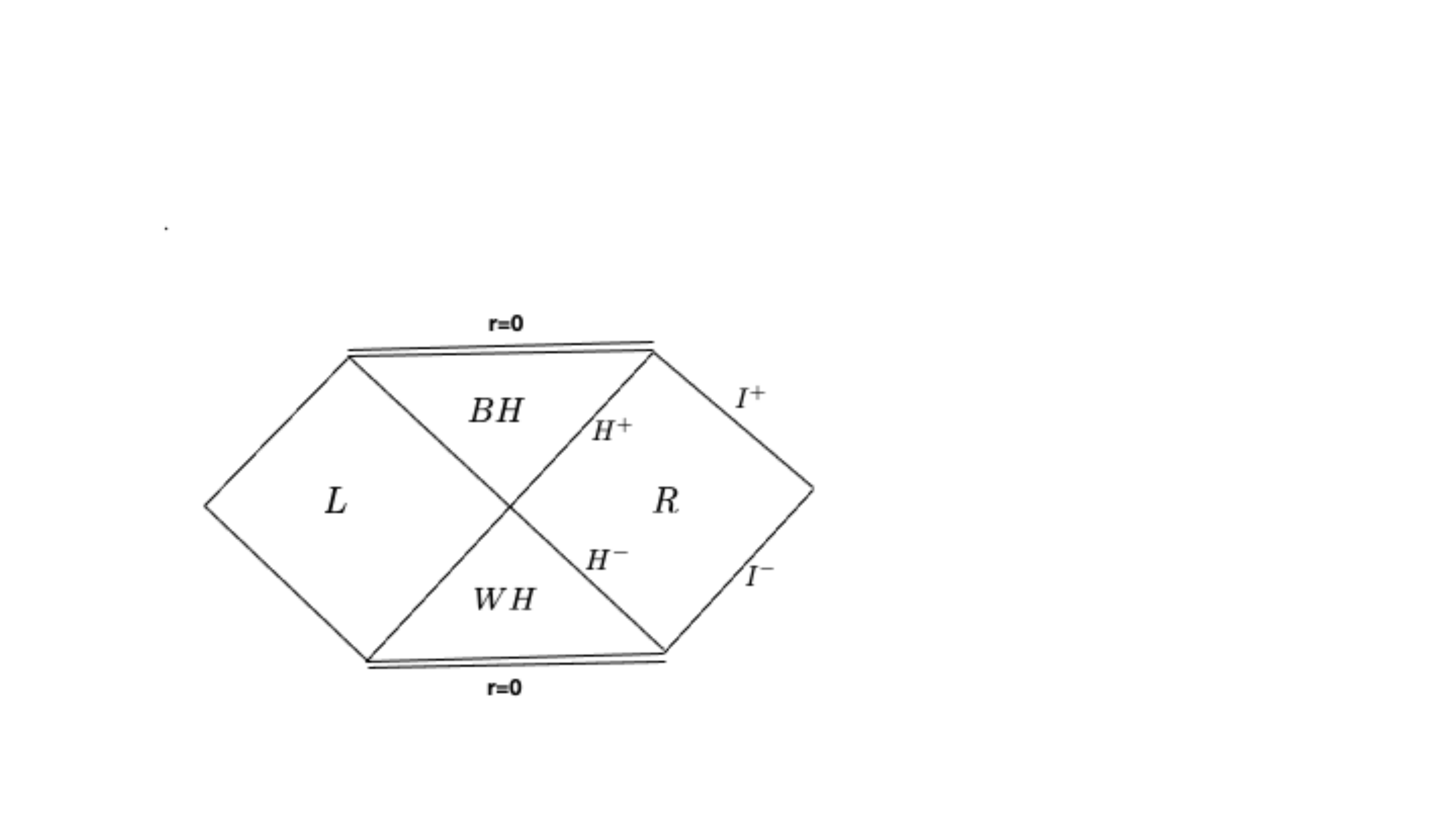}
\caption{Penrose diagram for the Schwarzschild geometry. $H^+$ is the future event horizon, $H^-$ the past one. $I^-$ is past null infinity, $I^+$ future null infinity.}
\label{figuno}
\end{figure}

The maximal analytic extension of the Schwarzschild metric is depicted in the Penrose diagram (see for example \cite{Hawking:1973uf}) of Fig.  (\ref{figuno}) describing the eternal Schwarzschild BH. In eq. (\ref{unoquattroa}) the sign $-$ holds in the asymptotically flat region $R$ and in the white hole region $WH$; the sign $+$ in the BH region $BH$ and in $L$. In (\ref{unoquattrob}) the sign $+$ refers to $R$ and $BH$ regions, while $-$ in $WH$ and $L$. We shall limit our discussion to the physically interesting regions $R$ and $BH$.

In this spacetime we consider a massless scalar quantum field $\hat\phi$ minimally coupled to gravity, whose field equation is
\be \label{unocinque}
\nabla_\mu\nabla^\mu \hat \phi =0\ , \ee
where $\nabla_\mu\nabla^\mu$ is the covariant D'Alembertian.

To get the Unruh vacuum one expands the field as 
\be \label{unosei}  \hat\phi = \sum_\omega \hat a_\omega^I \frac{e^{-i\omega v}}{\sqrt{4\pi\omega}} + \sum_{\omega_K} \hat a_{\omega_K} \frac{e^{-i\omega_K U}}{\sqrt{4\pi\omega_K}} + h.c. \ . \ee
The first term in eq. (\ref{unosei}) represents the ingoing part, while the second the outgoing one. Note that $U$ is locally inertial on $H^-$. 
The $\hat a$'s are annihilation operators and the Unruh vacuum $|U\rangle$ is defined as
\be \label{unosette} \hat a_\omega^I |U\rangle = 0 = \hat a_{\omega_K} |U\rangle \ee
for every $\omega, \omega_K$.

\begin{figure}[h]
\centering \includegraphics[angle=0, height=3.5in] {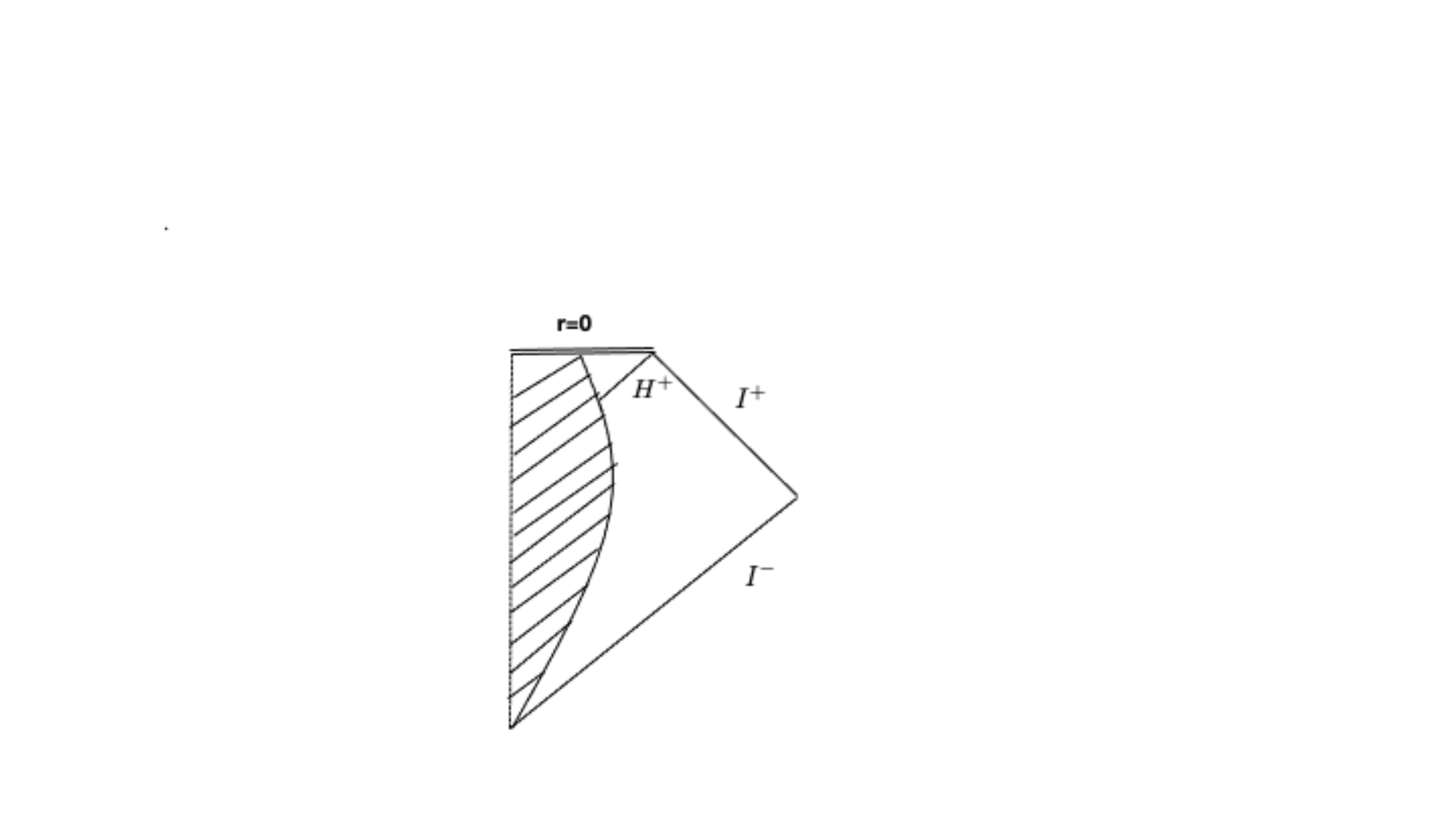}
\caption{Penrose diagram of the spacetime of a collapsing star The shadowed region represents the star.}
\label{figdue}
\end{figure}

Now consider a Schwarzschild BH formed by the collapse of a spherically symmetric star. The corresponding Penrose diagram is depicted in Fig. (\ref{figdue}). 
In this spacetime we consider, as before, a quantum field $\hat\phi$ and suppose the field is in a quantum state that corresponds to Minkowski vacuum on $I^-$. We call this state $|in\rangle$; in this state there are no incoming particles (radiation) associated to $\hat\phi$. 

Now $|U\rangle$ in the spacetime of Fig. (\ref{figuno}) approximates the behaviour of $|in\rangle$ in the physical spacetime of Fig. (\ref{figdue}) at late retarded time $u$. For a detailed discussion see \cite{Louko2}. In particular, considering the energy momentum tensor operator $\hat T_{\mu\nu}(\hat\phi)$ we have, in the Schwarzschild region,
\be \label{unootto} \lim_{u\to+\infty} \langle in|\hat T_{\mu\nu}(\hat\phi)|in\rangle = \lim_{u\to+\infty} \langle U|\hat T_{\mu\nu}(\hat\phi)|U\rangle \ . \ee
As it is well known, $|U\rangle$ describes at $I^+$ a thermal flux of radiation at the Hawking temperature
\be \label{unonove} T_H=\frac{\hbar\kappa}{2\pi k_B} \ , \ee
which is exactly what Hawking derived in 1974 \cite{hawking} computing the in-out Bogoliubov coefficients for the late time behaviour of $|in\rangle$. This is the reason for using the Unruh vacuum to describe Hawking's BH radiation.

In this paper we shall investigate the relation between the Unruh vacuum and the $|in\rangle$ vacuum in the setting of a charged BH described by the Reissner-Nordstr\"om (RN) metric, both for non extremal and extremal configurations. 

\section{Non extremal Reissner-Nordstr\"om black holes: the Unruh vacuum}
\label{s2}

\begin{figure}[h]
\centering \includegraphics[angle=0, height=2.5in] {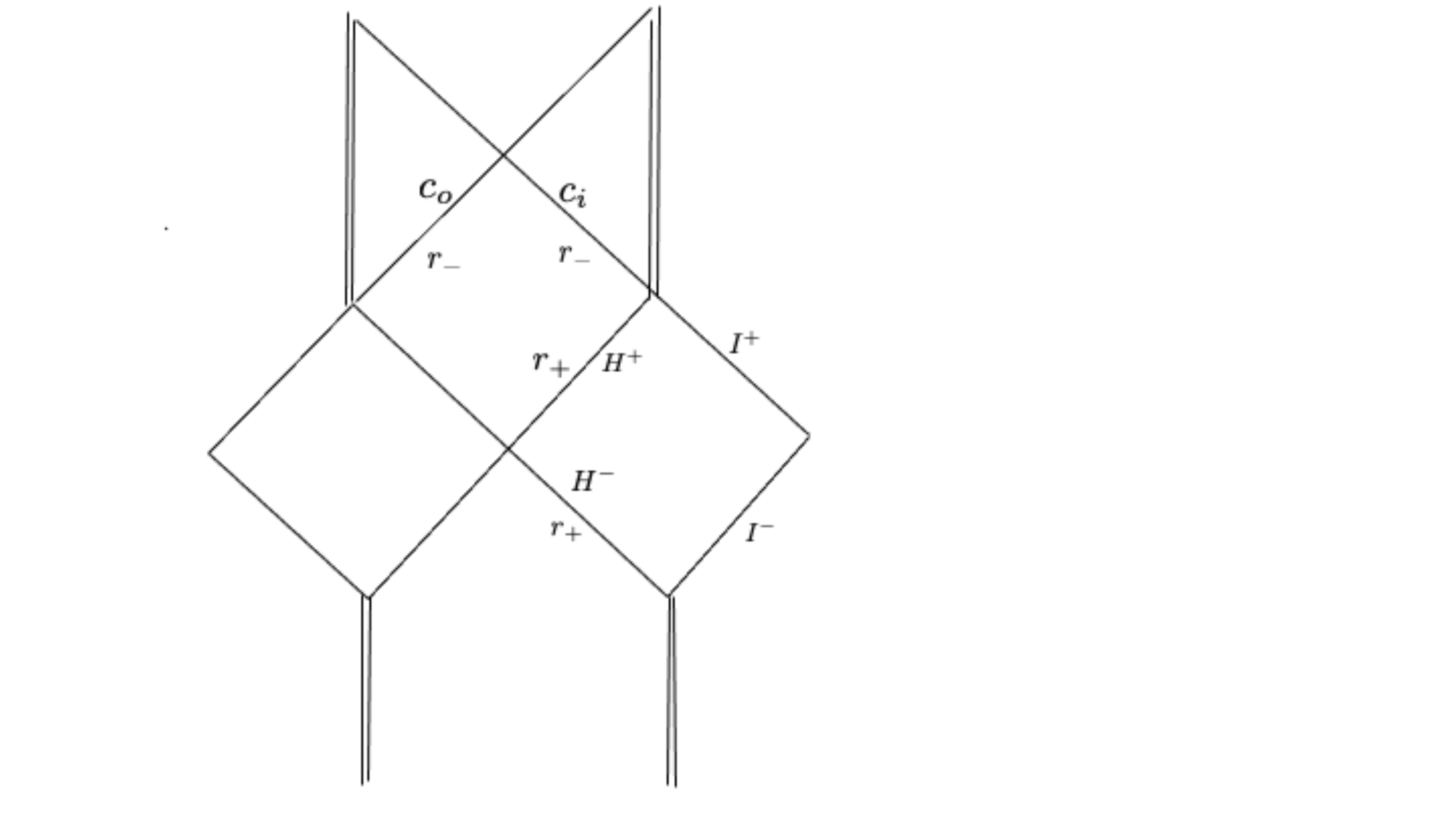}
\caption{Penrose diagram of the maximal analytic extension of the RN metric. $H^+$ is the future event horizon, $H^-$ the past one; $c_o$ and $c_i$ are, respectively, the outgoing and ingoing sheets of the Cauchy horizon located at $r_-$.}
\label{figtre}
\end{figure}

We  focus our attention on a RN BH of mass $m$ and charge $Q$. The BH is supposed to be non extremal, i,e, $m>|Q|$. The metric describing it is
\be \label{dueuno}
ds^2=-(1-\frac{2m}{r}+\frac{Q^2}{r^2})dt^2+(1-\frac{2m}{r}+\frac{Q^2}{r^2})^{-1}dr^2=-
(1-\frac{2m}{r}+\frac{Q^2}{r^2})dudv \ ,
\ee where again 
\bea
u=t-r_*\ , \label{dueduea} \\
v=t+r_*\ , \label{duedueb} \eea
but now 
\be r^*=\int \frac{dr}{(1-\frac{2m}{r}+\frac{Q^2}{r^2})}=r+\frac{1}{2\kappa_+}\ln|\kappa_+(r-r_+)|-\frac{1}{2\kappa_-}\ln|\kappa_-(r-r_-)|\ . \label{duetre} \ee
In this case we have two horizons, an event horizon at $r_+$ and an inner one at $r_-$ where
\be r_\pm = m\pm \sqrt{m^2-Q^2} \label{duequattro} \ee
and two corresponding surface gravities 
\be \kappa_\pm = \frac{\sqrt{m^2-Q^2}}{r_\pm^2}\ . \label{duecinque} \ee
We have two possible ways to implement the construction of Kruskal coordinates, namely 
\bea \label{dueseia}
U_{(+)}=\mp \frac{1}{\kappa_+}e^{-\kappa_+u}\ , \\
V_{(+)}=\frac{1}{\kappa_+}e^{\kappa_+v}\ , \label{dueseib}
\eea
which are regular on $r_+$ but not on $r_-$ where they diverge, or \cite{Poisson}
\bea \label{duesettea}
U_{(-)}= -\frac{1}{\kappa_-}e^{\kappa_-u}\ , \\
V_{(-)}=-\frac{1}{\kappa_-}e^{-\kappa_-v}\ , \label{duesetteb}
\eea
regular on $r_-$ but singular on $r_+$ where they diverge. The Penrose diagram of the maximally extended RN solution is given in Fig. (\ref{figtre}).
In the regions of our interest we have indicated as $H^+$ the future event horizon ($U_{(+)}=0, u=+\infty, r=r_+$) and $H^-$ the past one ($V_{(+)}=0, v=-\infty, r=r_+$). The Cauchy horizon $C=c_0\cup c_i$ is located at $r=r_-$: $c_o$ is the outgoing sheet ($U_{(-)}=0, u=-\infty, r=r_-$) while $c_i$ is the ingoing one ($V_{(-)}=0, v=+\infty, r=r_-$).

The Unruh vacuum is usually constructed out of $U_{(+)}$, namely expanding the field as 
\be\label{dueotto} \hat\phi=\sum_\omega\  \hat a_\omega^I \frac{e^{-i\omega v}}{\sqrt{4\pi\omega}} + \sum_{\omega_K}\ \hat a_{\omega_K} \frac{e^{-i\omega_K U_{(+)}}}{\sqrt{4\pi\omega_K}}+h.c.\ee
and the Unruh vacuum $|U_{(+)}\rangle$ is defined by 
\be \label{duenove}  \hat a_\omega^I  |U_{(+)}\rangle = 0 = \hat a_{\omega_K} |U_{(+)}\rangle \ee
for every $\omega, \omega_K$. 

We shall focus on the renormalized expectation values of $\hat T_{\mu\nu}(\hat\phi)$ 
\cite{Davies:1976ei, Birrel-davies, Fabbri-Navarro Salas} in this state which read \cite{Balbinot:2023grl} 
\bea  \langle U_{(+)}| \hat \hat T_{vv}(\hat\phi) |U_{(+)}\rangle &=& -\frac{1}{192\pi}\left[ f'(r)^2-2f(r)f''(r)\right] \nonumber \\ \label{duediecia} &=&\frac{1}{24\pi}\left(-\frac{m}{r^3}+\frac{3}{2} \frac{(m^2+Q^2)}{r^4} -\frac{mQ^2}{r^5}+\frac{Q^4}{r^6}\right)    \\ 
 \label{duediecib} \langle U_{(+)}| \hat \hat T_{uv}(\hat\phi) |U_{(+)}\rangle &=& \frac{1}{96\pi}f(r)f''(r)  = -\frac{1}{24\pi}\left(1-\frac{2m}{r}+\frac{Q^2}{r^2}\right)\left( \frac{m}{r^3}-\frac{3}{2} \frac{(m^2+Q^2)}{r^4} \right)\ \ \ \ \\
 \langle U_{(+)}| \hat \hat T_{uu}(\hat\phi) |U_{(+)}\rangle &=&  \langle U_{(+)}| \hat \hat T_{vv}(\hat\phi) |U_{(+)}\rangle-\frac{1}{24\pi} \{ U_{(+)},u\}  \nonumber \\ \label{duediecic} &=&\frac{1}{48\pi}\frac{(r-r_+)^2}{r^2} \left(\kappa_+^2(1+\frac{2r_+}{r}+ \frac{3r_+^2}{r^2})+ \frac{(\frac{r_-^2}{r_+}-3r_-)}{r^3} 
 +\frac{2r_-^2}{r^4}\right)\ ,  \ \ \ \eea
where 
\be \label{dueundici} f(r)=1-\frac{2m}{r}+\frac{Q^2}{r^2}\ee
and a prime ``$\ ' \ $'' indicates the derivative with respect to $r$. $\{ \ , \ \}$ means Schwarzian derivative, and we get
\be \label{duedodici}
-\frac{1}{24\pi} \{ U_{(+)},u \} =\frac{\kappa_+^2}{48\pi}\ . \ee
This last term in eq. (\ref{duediecic}) describes the Hawking radiation at 
\be \label{duetredici}
T_H=\frac{\hbar\kappa_+}{2\pi k_B}\ . \ee
The other terms in eqs.(\ref{duediecia})-(\ref{duediecic}) represent the vacuum polarization contribution and correspond to the expectation values calculated in the so called Boulware vacuum $|B\rangle$
\cite{Boulware},
obtained by expanding the field $\hat\phi$ in modes that are positive and negative frequency with respect to $t$, namely $( \frac{e^{-i\omega u}}{\sqrt{4\pi\omega}},\  \frac{e^{-i\omega v}}{\sqrt{4\pi\omega}} )$. We can formally write eqs. (\ref{duediecia})-(\ref{duediecic}) as
\be\label{duequattordici}  \langle U_{(+)}|\hat T_{\mu\nu}|U_{(+)}\rangle = \langle B |\hat T_{\mu\nu}| B \rangle + \Delta_{\mu\nu}^+\ , \ee
where
\be \label{duequindici} \Delta_{\mu\nu}^+=\langle U_{(+)}|\hat T_{\mu\nu}|U_{(+)}\rangle -  \langle B |\hat T_{\mu\nu}| B \rangle \ee
with
\bea && \Delta_{vv}^+=0=\Delta^+_{uv}\ , \label{duesedici} \\
&& \Delta_{uu}^+=-\frac{1}{24\pi}  \{ U_{(+)},u \} =\frac{\kappa_+^2}{48\pi}\ . \label{duediciassette} \eea
The tensor $\Delta_{\mu\nu}^+$ is conserved and describes radiation propagating along constant $u$.

Mathematically one can define also a Unruh vacuum associated to the Kruskal coordinate $U_{(-)}$
of eq. (\ref{duesettea}), i.e. expanding the field in outgoing modes of the form $\frac{e^{-i\omega_K U_{(-)}}}{\sqrt{4\pi\omega_K}}$. The resulting vacuum state, call it $|U_{(-)}\rangle$, has the following expectation values of $\hat T_{\mu\nu} $
\be\label{duediciotto}  \langle U_{(-)}|\hat T_{\mu\nu}|U_{(-)}\rangle = \langle B |\hat T_{\mu\nu}| B \rangle + \Delta_{\mu\nu}^-\ , \ee
where as before 
\be \label{duediciannove}
\Delta_{vv}^-=0=\Delta^-_{uv} \ , \ee
but now 
\be \label{dueventi} \Delta_{uu}^-=-\frac{1}{24\pi}  \{ U_{(-)},u \} =\frac{\kappa_-^2}{48\pi}\ .\ee

\section{Non extremal Reissner-Nordstr\"om black holes: the $|in\rangle$ vacuum}
\label{s3}

Let us now suppose that the RN BH is formed by gravitational collapse of a charged body. We simplify the discussion by considering the collapse of a charged null shell. One can generalize to an arbitrary body
\cite{FF} .

\begin{figure}[h]
\centering \includegraphics[angle=0, height=3in] {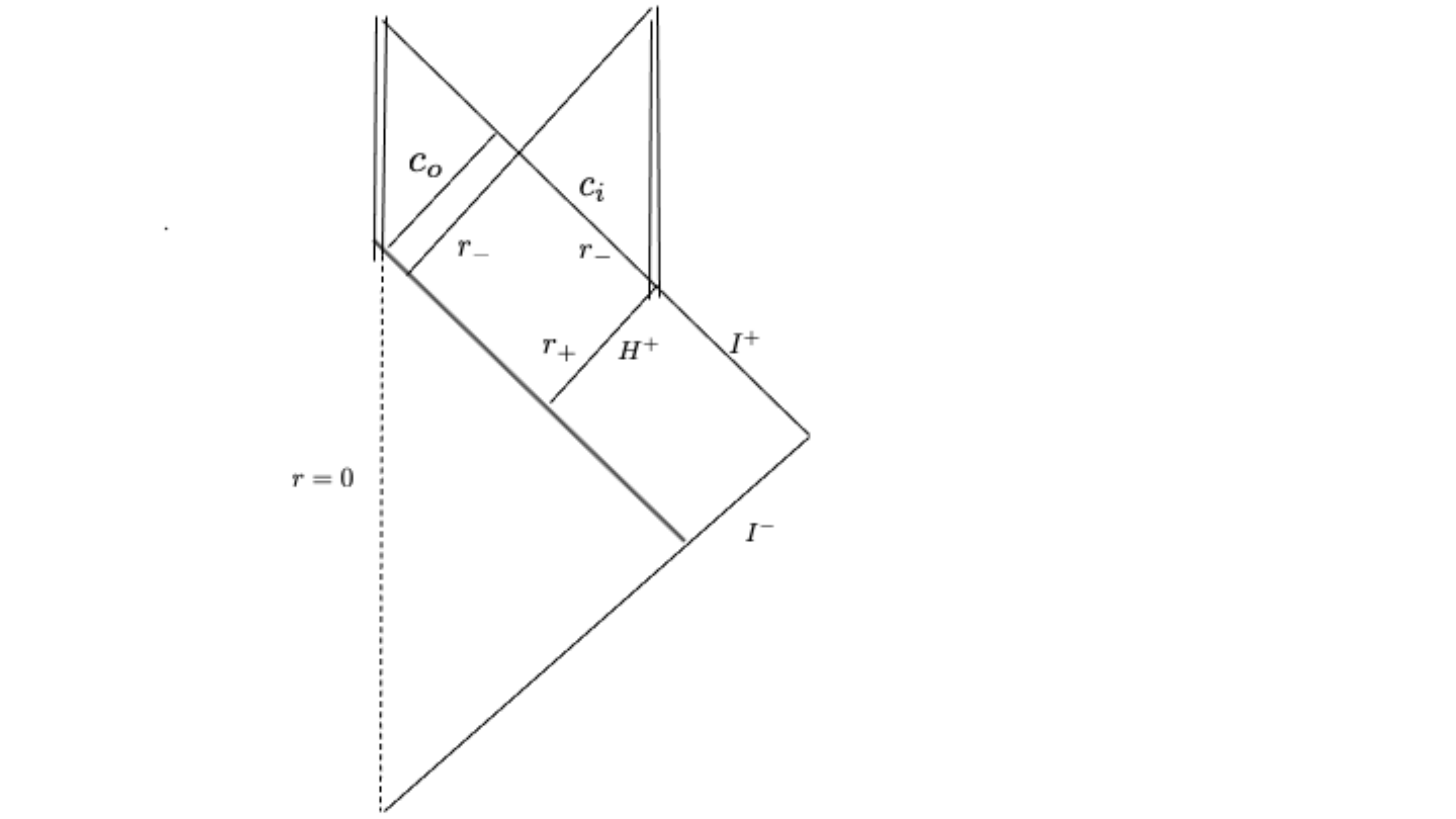}
\caption{Penrose diagram of the charged BH resulting from the collapse of a null shell.}
\label{figquattro}
\end{figure}

 The corresponding Penrose diagram is given in Fig. (\ref{figquattro}).

The spacetime metric of this model can be given as following. The shell is located at $v=v_0$. For $v<v_0$ we have Minkowski spacetime
\be\label{treuno} 
ds_{in}^2=-dt_{in}^2+dr^2=-du_{in}dv\ , \ee
where
\bea u_{in}=t_{in}-r\ , \label{treduea} \\
v=t_{in}+t\ , \label{tredueb} \eea
are, respectively, the null outgoing and ingoing directions. From eqs. (\ref{treduea}), (\ref{tredueb})
we have
\be \label{tretre} r=\frac{v-u_{in}}{2}\ . \ee
For $v>v_0$ we have the RN metric
\be \label{trequattro}
ds_{out}^2=-f(r)dt^2+f(r)^{-1}dr^2=-f(r)dudv\ , \ee
where $f(r)$ is given in eq. (\ref{dueundici}) and $u,v$ and $r_*$ in eqs. (\ref{dueduea}), (\ref{duedueb}) and (\ref{duetre}). Matching the metric across $v_0$ we can relate the null outgoing coordinates $u$ and $u_{in}$ \cite{Balbinot:2007kr}
\be \label{trecinque}
u=u_{in}-\frac{1}{\kappa_+}\ln|\kappa_+(v_0-u_{in}-2r_+)| +\frac{1}{\kappa_-}\ln|\kappa_-(v_0-u_{in}-2r_-)|\ . \ee
We can extend the $u_{in}$ coordinate in the RN region using the above relation. The event horizon 
$(r=r_+, u=+\infty)$ is located at 
\be \label{tresei} u_{in}|_{H^+}=v_0-2r_+\ . \ee
The inner horizon outgoing sheet $(r=r_-, u=-\infty)$ corresponds to 
\be \label{tresette} u_{in}|_{in}=v_0-2r_- \ . \ee
Note that this null surface, unlike the previous case of the maximal analytic extension of the RN spacetime (see Fig. (\ref{figtre})), does not correspond anymore to the outgoing sheet of the Cauchy horizon (see Fig. (\ref{figquattro})). This latter is indeed shifted inside the BH and is located at (using the reflection condition at $r=0$, see eq. (\ref{tretre})) 
\be \label{treotto}
u_{in}|_{c_o}=v_0\ . \ee

Note that near $H^+$ 
\be \label{trenove}
u\simeq -\frac{1}{\kappa_+}\ln |\kappa_+(v_0-u_{in}-2r_+)|\ , \ee
i.e. $u_{in}$ behaves as Kruskal $U_{(+)}$ of eq. (\ref{dueseia}) up to a shift. Choosing for example $v_0=2r_+$ they coincide.
On the other side, near the outgoing sheet of the inner horizon (i.e. for $u\to -\infty$) we have
\be \label{tredieci}
u\simeq \frac{1}{\kappa_-}\ln|\kappa_-(v_0-u_{in}-2r_-)|\ . \ee
This means that there $u_{in}$ behaves like $U_{(-)}$ (see eq. (\ref{duesettea})). So $u_{in}$, unlike $U_{(\pm)}$, is a regular null outgoing coordinate both at $(r_+,u=+\infty)$ and at $(r_-,u=-\infty$). 
However this coordinate
 is singular on the Cauchy horizon $c_o$, as can be seen from
\be ds^2=-C(r)dudv=-C(r)\frac{du}{du_{in}}du_{in}dv\ , \label{treundici} \ee
where from eq. (\ref{trecinque}) 
\be \label{tredodici}
\frac{du}{du_{in}}=1+\frac{1}{\kappa_+(v_0-u_{in}-2r_+)}-\frac{1}{\kappa_-(v_0-u_{in}-2r_-)}=\frac{(v_0-u_{in})^2}{(v_0-u_{in}-2r_+)(v_0-u_{in}-2r_-)} \ , \ee
which vanishes like $(u_{in}-v_0)^2$ for $u_{in}\to v_0$. Note that $C(r)\frac{du}{du_{in}}$ is nonvanishing neither on $H^+$ nor on the outgoing sheet of $r_-$ making the $(u_{in},v)$  coordinates regular there as stated before.

Note that past null infinity $I^-$ (see Fig (\ref{figquattro})) is a Cauchy surface for our null field for the spacetime region within the Cauchy horizon $c_o \cup c_i$. Initial conditions on our field $\hat\phi$ on $I^-$ determine its evolution in the above region. We choose the quantum state for our field to be Minkowski vacuum on $I^-$. This is achieved by choosing the ingoing modes to be of the form $e^{-i\omega v}$; the ougoing ones are simply the reflection through $r=0$ (in the Minkowski part) of the incoming ones, namely of the form $e^{-i\omega u_{in}}$ where the reflection condition sets $u_{in}=v$ (see eq. (\ref{tretre}) at $r=0$). We call this quantum state $|in\rangle$. 

Using the standard formalism we have 
\be \label{tretredici}
\langle in| \hat T_{\mu\nu} |in\rangle =0 \ee
for $v<v_0$ (i.e. in the Minkowski region), while for $v>v_0$ in the RN part we have 
\bea && \langle in | \hat T_{vv} |in\rangle = \langle B| \hat T_{vv} |B\rangle\ , \label{trequattordicia} \\
 && \langle in | \hat T_{uv} |in\rangle = \langle B| \hat T_{uv} |B\rangle\ , \label{trequattordicib} \\
&&  \langle in | \hat T_{uu} |in\rangle = \langle B| \hat T_{uu} |B\rangle -\frac{1}{24\pi} \{ u_{in},u \} \ , \label{trequattordicic}
 \eea
 where  the last term in eq. (\ref{trequattordicic}) is the Schwarzian derivative between $u_{in}$ and $u$ to be calculated from eq. (\ref{trecinque}) 
   \be \label{tresedici}
\{ u_{in}, u \} = \frac{3}{2}\kappa_+^2 \frac{\left( 1- \frac{\kappa_+}{\kappa_-} \frac{U_{in}^2}{(U_{in}-2(r_+-r_-))^2}\right)^2}{\left(1-\kappa_+ U_{in}-\frac{\kappa_+}{\kappa_-}\frac{U_{in}}{(U_{in}-2(r_+-r_-))}\right)^4}-2\kappa_+^2\frac{\left( 1- \frac{\kappa_+}{\kappa_-} \frac{U_{in}^3}{(U_{in}-2(r_+-r_-))^3}\right)}{\left(1-\kappa_+ U_{in}-\frac{\kappa_+}{\kappa_-}\frac{U_{in}}{(U_{in}-2(r_+-r_-))}\right)^3}\ , \ee
where $U_{in}=u_{in}-v_0+2r_+$. 
The following limiting behaviours are interesting. For $u\to +\infty$ (i.e. $u_{in}=v_0-2r_+, U_{in}= 0$) we have (see eq. (\ref{trenove}))
\be \label{trediciassettea} -\frac{1}{24\pi} \{ u_{in},u \}=-\frac{1}{24\pi} \{ U_{(+)},u \}  =\frac{1}{48\pi} \kappa_+^2\ , \ee
while for $u\to -\infty$ (i.e. $u_{in}=v_0-2r_-, U_{in}=2(r_+-r_-)$), see eq. (\ref{tredieci}), 
\be \label{trediciassetteb} -\frac{1}{24\pi} \{ u_{in},u \}=-\frac{1}{24\pi} \{ U_{(-)},u \}  =\frac{1}{48\pi} \kappa_-^2\ . \ee
So, comparing eqs. (\ref{trequattordicia})-(\ref{trequattordicic}) with the corresponding ones (\ref{duediecia})-(\ref{duediecic}) and (\ref{duediciotto}) we can conclude that in the RN region
\be \label{trediciottoa}
\lim_{u\to +\infty} \langle in| \hat T_{\mu\nu} |in\rangle = \lim_{u\to +\infty} \langle U_{(+)}| \hat T_{\mu\nu} | U_{(+)}\rangle \ee
and
\be \label{trediciottob}
\lim_{u\to -\infty} \langle in| \hat T_{\mu\nu} |in\rangle = \lim_{u\to -\infty} \langle U_{(-)}| \hat T_{\mu\nu} | U_{(-)}\rangle  \ .\ee
So the Unruh vacuum $|U_{(+)}\rangle$ reproduces $|in\rangle$ at late ($u\to +\infty$) retarded time, while this does not happen at the inner horizon. There the $|in\rangle$ vacuum is well approximated
by $|U_{(-)}\rangle$.

\section{Regularity}
\label{s4}

We shall consider a quantum state to be ``regular'' if the expectation values of $\hat T_{\mu\nu}$ in that given state are finite when expressed in a regular coordinate system.
Support to this comes from the fact that the expectation values of $\hat T_{\mu\nu}$
in the semiclassical Einstein equations represent the source which drives the backreaction of the quantum fields on the spacetime metric. 

Let us start with the Unruh vacuum $|U_{(+)}\rangle$. The expectation values of $\hat T_{\mu\nu}$
in this state are given by eqs. (\ref{duediecia})-(\ref{duediecic}) and one sees immediately that they diverge on the physical singularity located at $r=0$. However the Eddington-Finkelstein coordinates $(u,v)$ used in eqs. (\ref{duediecia})-(\ref{duediecic}) are singular on the horizons $r_{\pm}$. So care must be taken when considering these regions. 

We begin with $r_+$. A coordinate system regular there is the Kruskal one $(U_{(+)}, V_{(+)})$ of eqs, (\ref{dueseia}),(\ref{dueseib}). We have to require finiteness of $\langle U_{(+)}|\hat T_{\mu\nu}|U_{(+)}\rangle$ as $r\to r_+$ when expressed in these Kruskal coordinates. This implies that on the event horizon $H^+$ $(r=r_+, U_{(+)}=0)$ we need for the regularity of $|U_{(+)}\rangle$  (see \cite{Christensen:1977jc}) that
\bea && i)\ \ \ \ \frac{\langle U_{(+)}|\hat T_{uu}|U_{(+)}\rangle}{f^2}<\infty\ , \label{quattrounoa} \\
&& ii)\ \ \ \ \frac{\langle U_{(+)}|\hat T_{uv}|U_{(+)}\rangle}{f}<\infty\ , \label{quattrounob} \\
&& iii)\ \ \ \ \langle U_{(+)}|\hat T_{vv}|U_{(+)}\rangle <\infty\ . \label{quattrounoc}
 \eea
Conditions $ii)$ and $iii)$ are easily seen to be satisfied using eq. (\ref{duediecib}) and (\ref{duediecia})
respectively. Concerning condition $i)$ note that
\be \label{quattrodue} \lim_{r\to r_+} \langle B|\hat T_{uu} |B\rangle = - \frac{1}{48\pi} \kappa_+^2\ . \ee
So given eq. (\ref{duediciassette}) we have that $\langle U_{(+)}|\hat T_{uu}|U_{(+)}\rangle$ vanishes at $r=r_+$ and also, as can be rapidly checked, its first derivative. This means that $\langle U_{(+)}|\hat T_{uu}|U_{(+)}\rangle$ vanishes as $(r-r_+)^2$ for $r\to r_+$ making condition $i)$ satisfied.
So $|U_{(+)}\rangle$ is regular on $H^+$. 
Regularity on  $H^-$ is given by conditions similar to eqs. (\ref{quattrounoa})-(\ref{quattrounoc}) but with $u$ and $v$ interchanged, namely
\bea && i)\ \ \ \ \frac{\langle U_{(+)}|\hat T_{vv}|U_{(+)}\rangle}{f^2}<\infty\ , \label{quattrotrea} \\
&& ii)\ \ \ \ \frac{\langle U_{(+)}|\hat T_{uv}|U_{(+)}\rangle}{f}<\infty\ , \label{quattrotreb} \\
&& iii)\ \ \ \ \langle U_{(+)}|\hat T_{uu}|U_{(+)}\rangle <\infty\ . \label{quattrotrec}
 \eea
 From eqs. (\ref{duediecib}) and (\ref{duediecic}) we see that $ii)$ and $iii)$ are satisfied, but not $i)$. From eq. (\ref{duediecia}) we see that 
 \be \lim_{r\to r_+} \langle U_{(+)}|\hat T_{vv}|U_{(+)}\rangle = \lim_{r\to r_+} \langle B |\hat T_{vv}|B \rangle =-\frac{1}{48\pi}\kappa_+^2  \label{quattroquattro} \ee
and hence condition $i)$ is not satisfied as $f\to 0$. Hence $|U_{(+)}\rangle$ is not regular on $H^-$. 

We now come to the inner horizon, where a coordinate system regular there can be given by Kruskal's $(U_{(-)},V_{(-)})$, eqs. (\ref{duesettea}), (\ref{duesetteb}).   The regularity conditions on the outgoing sheet of the Cauchy horizon $(r=r_-, U_{(-)}=0)$ are given by the same requirements of eqs. (\ref{quattrounoa})-(\ref{quattrounoc}). One rapidly checks the fullfillement of eq. (\ref{quattrounob}) and (\ref{quattrounoc}). Concerning (\ref{quattrounoa}) we have now that
\be \label{quattroquattro} 
\lim_{r\to r_-} \langle B|\hat T_{uu}|B\rangle =-\frac{1}{48\pi}\kappa_-^2 \ee
implying that (see eq. (\ref{duediciassette})) 
\be \label{quattrocinque}
\lim_{r\to r_-} \langle U_{(+)}|\hat T_{uu}|U_{(+)}\rangle =-\frac{1}{48\pi}(\kappa_-^2-\kappa_+^2)\ . \ee
Being $\kappa_+\neq \kappa_-$, this is not vanishing and condition (\ref{quattrounoa}) is not fullfilled. $|U_{(+)}\rangle$ is not regular on $c_o$.  
The conditions for regularity on the ingoing sheet of the inner horizon $(r=r_-, V_{(-)}=0)$ are given by eqs. (\ref{quattrotrea})-(\ref{quattrotrec}) and one checks that eq. (\ref{quattrotrea}) is not satisfied for $f\to 0$ since
\be \lim_{r\to r_-} \langle U_{(+)}|\hat T_{vv}|U_{(+)}\rangle = \lim_{r\to r_-} \langle B |\hat T_{vv}|B \rangle = -\frac{1}{48\pi}\kappa_-^2 \ . \label{duattrosei} \ee
$|U_{(+)}\rangle$ is also singular on $c_i$ \cite{hiscock77, Birrell:1978th} (see also \cite{Louko1}). 

A similar analysis can be performed on the state $|U_{(-)}\rangle$ (see eqs. (\ref{duediciotto})-(\ref{dueventi})) and we will find that $|U_{(-)}\rangle$ is not regular on both future horizon $H^+$ and past horizon $H^-$ and also on the ingoing sheet of the inner horizon $c_i$, while it is regular on the outgoing sheet $c_o$.

We come now to the $|in\rangle$ state. In view of the previous discussion on the states $|U_{(\pm)}\rangle$ and the relations of these with the $|in\rangle$ state found in section \ref{s3}, we have the following behaviour. On $H^+$ $(r=r_+,u=+\infty)$ we have that $|in\rangle$ is regular, since (see (\ref{trediciottoa}))
\be \label{quattrosette}
\lim_{u\to +\infty} \langle in|\hat T_{\mu\nu} |in\rangle = \lim_{u\to +\infty} \langle U_{(+)}|\hat T_{\mu\nu} |U_{(+)} \rangle \ee
and $|U_{(+)}\rangle$ is regular on $H^+$. On the inner horizon $(r=r_-, u=-\infty)$, which is no longer the outgoing sheet of the Cauchy horizon, from (see eq. (\ref{trediciottob}))
\be \label{quattrootto}
\lim_{u\to -\infty} \langle in|\hat T_{\mu\nu} | in\rangle = \lim_{u\to -\infty} \langle U_{(-)}|\hat T_{\mu\nu}|U_{(-)}\rangle  \ee
we have that $|in\rangle$, unlike $|U_{(+)}\rangle$, is regular because $|U_{(-)}\rangle$ is so. 
Furthermore the non regularity of $|in\rangle$ on the past horizon $H^-$ $(r=r_+, v=-\infty)$ and on the ingoing sheet of the inner horizon $c_i$ $(r=r_-, v=+\infty)$ is rapidly established since
\be \label{quattronove} \langle in|\hat T_{vv}|in\rangle=  \langle U_{(+)} |\hat T_{vv}|U_{(+)} \rangle 
= \langle U_{(-)} |\hat T_{vv}|U_{(-)} \rangle =\langle B |\hat T_{vv}| B \rangle\ ,  \ee 
and we have seen that $|U_{(\pm)}\rangle$ are not regular on these regions. As we have remarked presenting the spacetime of the collapsing shell (see Fig. (\ref{figquattro})), the outgoing sheet of the Cauchy horizon is no longer located at $(r=r_-,u=-\infty)$ but at $u_{in}=v_0$. To check regularity on this surface we need just to examine the component of $\langle in|\hat T_{\mu\nu}|in\rangle$ in Eddington-Finkelstein coordinates $(u,v)$ since they are regular there. Examining eqs. (\ref{trequattordicia})-(\ref{trequattordicic}) we find that the Schwarzian derivative $\{ u_{in},v\}$ entering in (\ref{trequattordicic}) diverges as $(u_{in}-v_0)^{-6}$, all other terms in eqs. (\ref{trequattordicia})-(\ref{trequattordicic}) are finite (except at the singularity at $r=0$). We conclude that $|in\rangle$ is not regular on $c_o$.

\section{Extremal Reissner-Nordstr\"om BHs}
\label{s5}

So far, we have considered non extremal BHs, i.e. $m>|Q|$, but what happens if the BH is extremal? 
Extremal BHs are very interesting objects: they appear also in supergravity theories \cite{ADFT} and represent in some sense a stable ground state for these theories.  For extremal BHs the surface gravity vanishes (see eq. (\ref{duecinque}) for $m^2=Q^2$) and so they do not emit Hawking radiation. They can be considered as the end state of the evaporation of a non extremal BH when the $U(1)$ charge $Q$ is conserved during the process. 

\begin{figure}[h]
\centering \includegraphics[angle=0, height=2.5in] {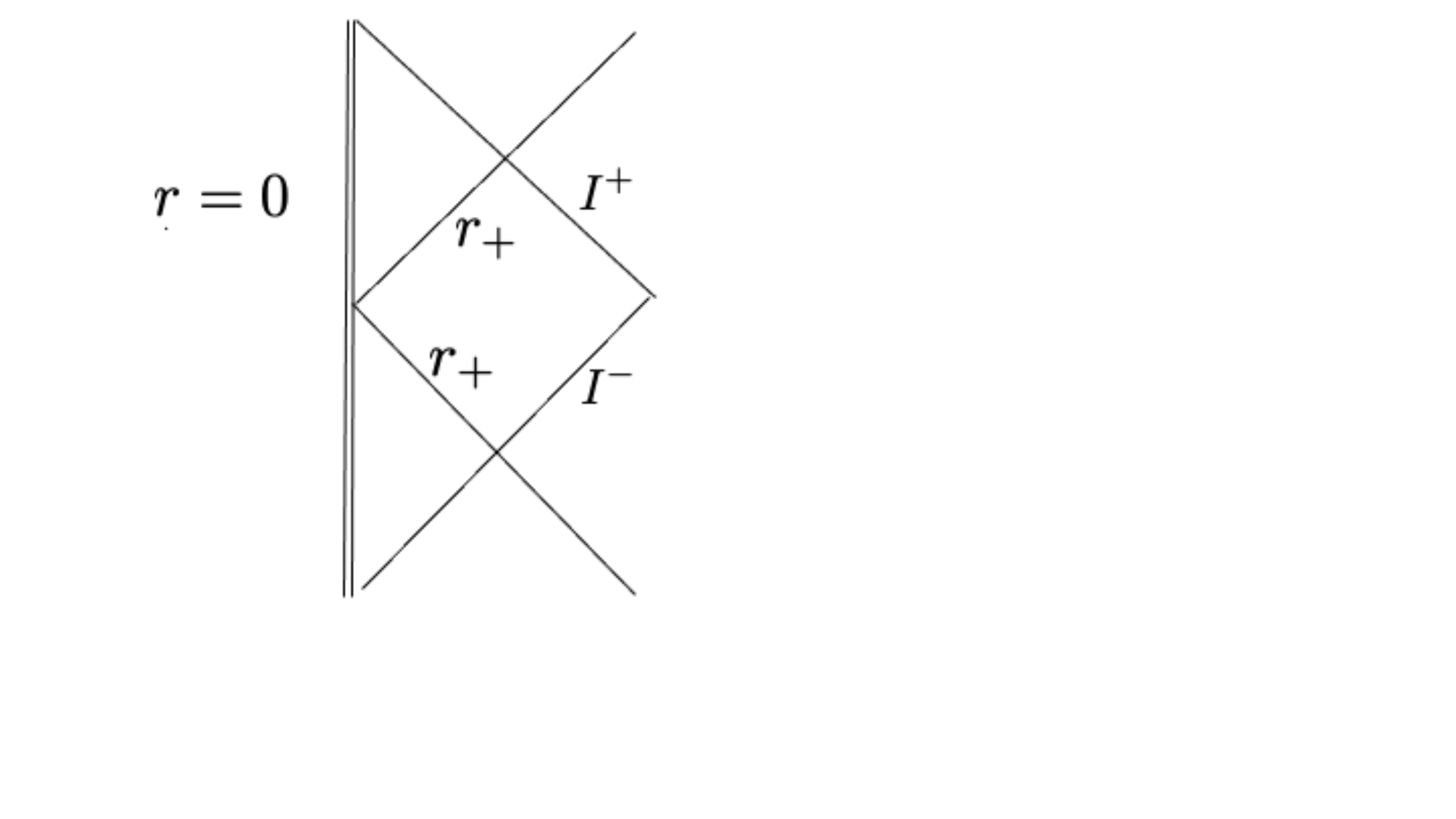}
\caption{Penrose diagram of the extremal RN BH.}
\label{figcinque}
\end{figure}

We shall now discuss how we can extend the results of our previous investigation to the extremal BH case. The extremal RN BH is characterized by a metric
\be \label{cinqueuno} ds^2=-\left(1-\frac{m}{r}\right)^2dt^2 +\left(1-\frac{m}{r}\right)^{-2}dr^2=-\left(1-\frac{m}{r}\right)^2dudv\ , \ee
where as usual
\bea \label{cinqueduea} && u=t-r^*\ ,  \\ && v=t+r^* \ , \label{cinquedueb} \eea
but now
\be \label{cinquetre} r^*=\int \frac{dr}{(1-\frac{m}{r})^2}=r+2m \ln | \frac{r}{m}-1| -\frac{m^2}{(r-m)^2}\ .\ee
Note the presence of the last term in (\ref{cinquetre}) compared to the non extremal case. The Penrose diagram of the extremal RN BH is given in Fig. (\ref{figcinque}).
Inner and outer horizon now coincide at $r_+=m$.

We start by considering the Unruh vacuum. If we simply look at the final expression for the expectation values of $\hat T_{\mu\nu}$ in the state $|U_{(+)}\rangle$ (eqs. (\ref{duediecia})-(\ref{duediciassette})) and naively set $m=|Q|$, hence $\kappa_+=0$, we see that the Schwarzian derivative term (\ref{duedodici}) vanishes, so the Unruh state coincides with the Boulware one leading to 
\bea \label{cinquecinquea} \langle B| \hat T_{vv} |B\rangle &=& -\frac{1}{24\pi} \frac{m}{r^3}\left(1-\frac{m}{r}\right)^3\ , \\
\label{cinquecinqueb} \langle B| \hat T_{uv} |B\rangle &=& -\frac{1}{24\pi} \left(1-\frac{m}{r}\right)^2\left(\frac{m}{r^3}-\frac{3}{2}\frac{m^2}{r^4}\right) \ , \\ 
\label{cinquecinquec} \langle B| \hat T_{uu} |B\rangle &=& -\frac{1}{24\pi} \frac{m}{r^3}\left(1-\frac{m}{r}\right)^3\ , 
\eea
which describes just vacuum polarization induced by the BH and this seems consistent with the fact that extremal BHs do not Hawking radiate. 

It is easy to see that $\langle B|\hat T_{\mu\nu}|B\rangle$ is not regular on the horizon $r_+=m$ \cite{Trivedi:1992vh}. Requiring regularity in the free falling frame across the future horizon we get conditions identical to eqs. 
(\ref{quattrounoa})-(\ref{quattrounoc}) and we see that (\ref{quattrounoa}) is not satisfied, since now the denominator there vanishes as $(r-m)^4$ while the numerator only as $(r-m)^3$ (see eq. (\ref{cinquecinquec})). Similarly, one finds a singular behaviour on the past horizon and on the Cauchy horizon. It is not yet clear if this singular behaviour is an artifact of the two-dimensional theory
\cite{AHL, Arrechea:2023fas}.

Beside regularity issues, one can question if the limiting procedure we used to derive eqs. (\ref{cinquecinquea})-(\ref{cinquecinquec}) as the $\kappa_+\to 0$ limit of the non extremal ones (\ref{duediecia})-(\ref{duediecic}) is correct. If we go back to the original definition of the Unruh state $|U_{(+)}\rangle$ starting with outgoing modes that are positive frequency with respect to Kruskal $U_{(+)}$, we see that the equation (\ref{dueseia}) defining $U_{(+)}$ in terms of Eddington-Finkelstein $u$ makes no sense if $\kappa_+=0$. So one may wonder if the stress tensor of eqs. (\ref{cinquecinquea})-(\ref{cinquecinquec}) really describes the behaviour of an extremal RN BH at late retarded time after its formation, as indeed it happens for a non extremal BH as we saw in section \ref{s3}. To answer this one should investigate the formation of an extremal RN BH, then construct the corresponding $|in\rangle$ state, compute the expectation values of $\hat T_{\mu\nu}$ in this state and see if at late $u$ they coincide with the predictions of eqs. (\ref{cinquecinquea})-(\ref{cinquecinquec}). 

\begin{figure}[h]
\centering \includegraphics[angle=0, height=3in] {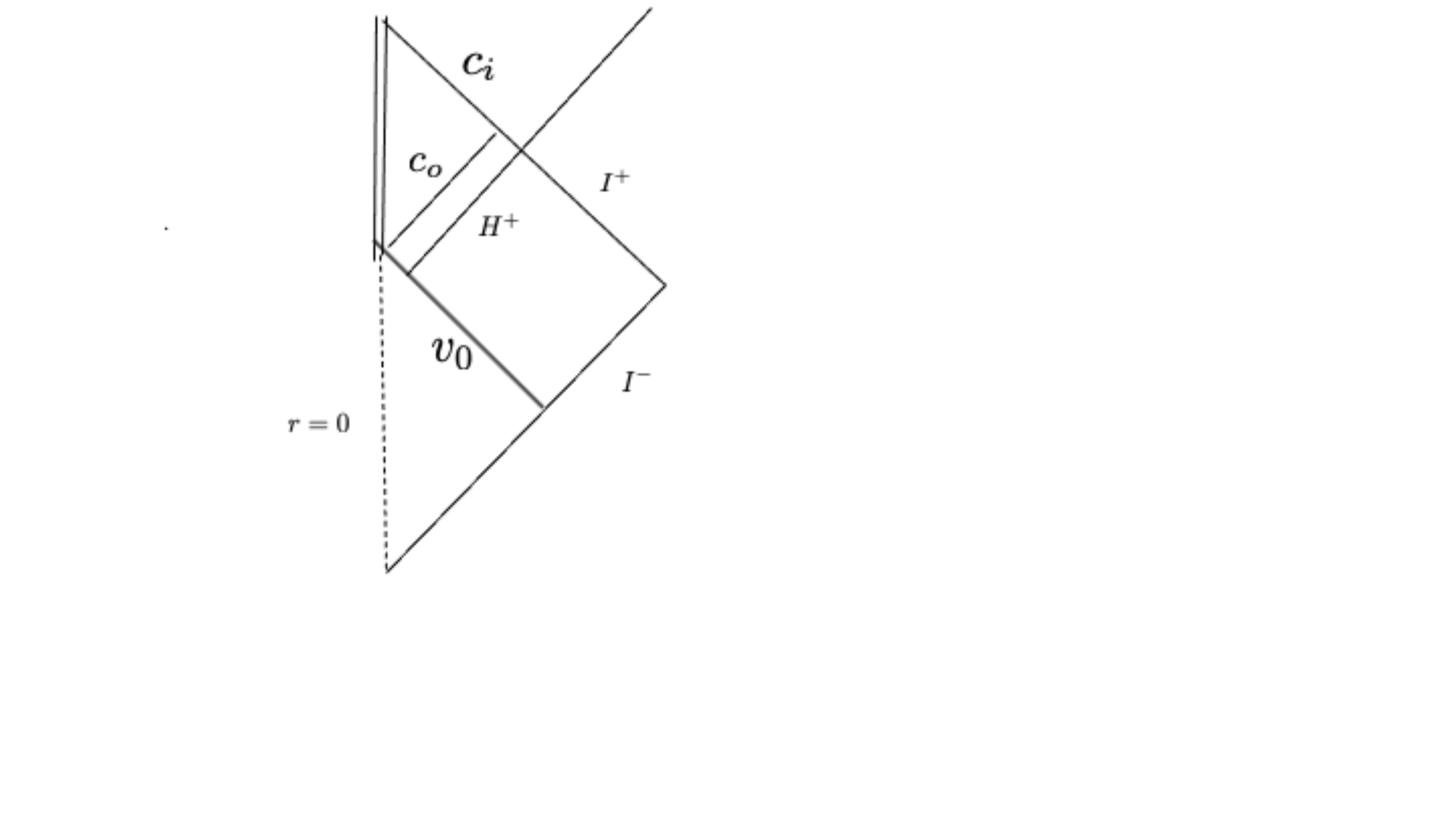}
\caption{Penrose diagram of an extremal RN BH formed by the collapse of a null shell.}
\label{figsei}
\end{figure}

Consider therefore the formation of an extremal RN BH, as we did before, by the collapse of a charged null shell. One could consider instead a generic collapse without changing our conclusions. The Penrose diagram is given in Fig. (\ref{figsei}). 
For $v<v_0$ we have Minkowski spacetime
\bea \label{cinquesette} && ds^2_{in}=-dt_{in}^2+dr^2=-du_{in}dv\ , \\
\label{cinqueottoa} && u_{in}=t_{in}-r\ , \\
\label{cinqueottob} && v=t_{in}+r\ , \eea
while for $v>v_0$ the extreme RN metric given by eq. (\ref{cinqueuno}). Matching across the shell located at $v=v_0$ gives 
\be \label{cinquenove} u=u_{in}-4m \left[ \ln|\frac{v_0-u_{in}}{2m}-1| - \frac{1}{2(\frac{v_0-u_{in}}{2m}-1)} \right] \ . \ee
The event horizon $H^+$ $(r=m,u=+\infty)$ is at $u_{in}=v_0-2m$. The ingoing sheet of the Cauchy horizon $c_i$ is at $(r=m,v=+\infty)$, while the outgoing one $c_o$ is at $u_{in}=v_0$. Using eq. (\ref{cinquenove}) one can calculate  the Schwarzian derivative and the complete stress tensor reads
\bea \label{cinquediecia} \langle in| \hat T_{vv} |in\rangle &=& \langle B| \hat T_{vv} |B\rangle=-\frac{1}{24\pi} \frac{m}{r^3}\left(1-\frac{m}{r}\right)^3\ , \\
\label{cinquediecib} \langle in| \hat T_{uv} |in\rangle &=& \langle B| \hat T_{uv} |B\rangle= -\frac{1}{24\pi} \left(1-\frac{m}{r}\right)^2\left(\frac{m}{r^3}-\frac{3}{2}\frac{m^2}{r^4}\right) \ , \\ 
\label{cinquediecic} \langle in| \hat T_{uu} |in\rangle &=& \langle B| \hat T_{uu} |B\rangle -\frac{1}{24\pi}\{ u_{in},u \} \\ &=& -\frac{1}{24\pi}\frac{m}{r^3}\left(1-\frac{m}{r}\right)^3  - \frac{1}{24\pi}\frac{8m\left(u_{in}-v_0+2m\right)^3}{\left(u_{in}-v_0\right)^6} \ . \nonumber
\eea
Looking at the $(u,u)$ component, we immediately see that it has an extra contribution with respect to the static vacuum polarization $ \langle B| \hat T_{uu} |B\rangle$ of eq. (\ref{cinquecinquec}). This term represents transient radiation created by the time dependent collapse of the shell decaying at late retarded time $(u\to +\infty, u_{in}=v_0-2m)$ as $(u_{in}-v_0+2m)^3$. But near the horizon we have $u_{in}-v_0+2m\sim -2(r-m)$, so the two terms in eq. (\ref{cinquediecic}) vanish with the same power law and they are opposite in sign. This makes the crucial limit for the regularity condition on $H^+$, i.e. eq. (\ref{quattrounoa}),  satisfied \cite{Balbinot:2004jx} 
\be \label{cinqueundici} \frac{\langle in | \hat T_{uu} |in\rangle}{(1-\frac{r}{m})^4}=-\frac{1}{24\pi}(\frac{3}{2m^2}) <\infty\ . \ee
So $|in\rangle$ is regular on $H^+$. We conclude that $|B\rangle$ cannot represent the state of a quantum field emerging at late retarded time for a collapse forming an extremal RN BH.

Note that in order to get the nonvanishing term coming from the Schwarzian derivative it is necessary to keep also the subleading logarithmic term in eq. (\ref{cinquenove}) in the limit $u_{in}\to v_0-2m$. Omission of it would result in a vanishing Schwarzian derivative, missing therefore the correct result.\footnote{An expression like eq. (\ref{cinquenove}) with the first term omitted has been proposed in Ref. \cite{Liberati:2000sq, gao} as definition of Kruskal $U$ coordinate in the extreme RN metric. The vacuum state constructed out of it would approximate $|in\rangle$ quite well for $u\to +\infty$.}

Finally, one can verify that $|in\rangle$ is not regular on the ingoing sheet of the Cauchy horizon and also on the outgoing one $(u_{in}=v_0)$, where the Schwarzian derivative diverges like $(u_{in}-v_0)^{-6}$. 

\section{Conclusion}

We have investigated two particularly interesting quantum states for a field that propagates in the RN spacetime: the Unruh state $|U_{(+)}\rangle$ and the $|in\rangle$ state. They are defined in two different spacetime manifolds. The Unruh state $|U_{(+)}\rangle$ is defined on the maximal analytic extension of the RN metric (see Fig. (\ref{figtre})). It is a mathematical manifold. The $|in\rangle$ state is defined on the manifold associated to the formation of a RN BH (see Fig. (\ref{figquattro})). This is the physical manifold. We have seen if and when the Unruh state $|U_{(+)}\rangle$ can mimic the behaviour of the physical state $|in\rangle$.

These two states differ by the choice of outgoing modes in the expansion of the field operator, while the ingoing ones are identical. For the Unruh vacuum, for a non extremal BH the outgoing modes are positive frequency with respect to Kruskal's null retarded coordinate $U_{(+)}$ (see eq. (\ref{dueseia})). This coordinate is locally inertial on the past horizon $H^-$. On the other hand, for the $|in\rangle$ vacuum the outgoing modes are simply the reflection of the ingoing ones at the center of the collapsing body. The incoming modes in this case are chosen in a way that the corresponding vacuum $|in\rangle$  coincides, at past null infinity $I^-$ (which is a Cauchy surface), with Minkowski vacuum. One can see that the definition of the $|in\rangle$ vacuum has a very solid physical motivation, it is not just a mathematical construction. 

We have seen that for a non extremal BH $|U_{(+)}\rangle$ approximates very well $|in\rangle$ at late retarded time ($u\to +\infty$): this means near the horizon $H^+$ and asymptotically at late-time. This motivates the use of the Unruh vacuum to discuss BH evaporation. However, although it is certainly  true that the agreement between the two states is quite good near the horizon $H^+$, this is no more true as soon as one moves inside the BH. While both states are regular on the event horizon $H^+$, the Unruh vacuum is singular on the outgoing sheet of the inner horizon at $r_-$ for the maximal analytic extension of the RN metric. The state $|in\rangle$, on the other hand, is perfectly regular on this outgoing null surface which is no longer a Cauchy horizon for the collapse spacetime manifold of Fig. (\ref{figquattro}). For this physical spacetime the outgoing sheet of the Cauchy horizon is shifted inside the inner horizon $r_-$ at $u_{in}=v_0$, where $|in\rangle$ is singular. Note that in the maximal analytic extension this surface is completely regular and has no particular physical significance. From eq. (\ref{trequattordicic}) we see that it is the Schwarzian derivative term which for the state $|in\rangle$ is diverging as $u_{in}\to v_0$. This term represents the radiation emitted by the shell during its collapse. We see that this flux is diverging as the shell radius goes to zero: we have a thunderbolt null singularity
\cite{CGHS, HS}. 

Concerning the inner sheet of the Cauchy horizon, located for both spacetimes at $(r=r_-, v=+\infty)$, we have that $|U_{(+)}\rangle$ and $|in\rangle$ are both not regular making the full Cauchy horizon singular. One expects large backreaction effects through the semiclassical Einstein equations to occur there leading to the formation of a spacetime singularity as predicted by the mass inflation mechanism \cite{Poisson:1989zz}, \cite{Balbinot:1993rf}. 

We finally analysed the case of an extreme BH. It is often said that the Unruh vacuum (and also the Hartle-Hawking-Israel one \cite{Hartle:1976tp,Israel:1976ur}) coincide with the Boulware one since the surface gravity of these BHs horizons is vanishing. One may further support this idea by remarking the fact that the Unruh vacuum is constructed by outgoing modes which are positive frequency with respect to a null coordinate ($U_{(+)}$ for a non extremal BH) which is locally inertial on the past horizon. For an extremal RN BH, Eddington-Finkelstein $u$ is locally inertial on the past horizon and the outgoing modes of the Boulware vacuum are indeed positive frequency with respect to $u$. However, we have clearly shown that this Boulware vacuum for an extremal BH does not correctly describe at large $u$ the physical state of a quantum field if the extreme RN BH is formed by the collapse of a star. So its physical significance is quite unclear and the results for extremal BHs obtained by simply taking the $\kappa\to 0$ limit of expressions calculated for the Unruh or Hartle-Hawking-Israel state in non extremal BHs should be regarded with suspicion. 

 \acknowledgments
 
A.F. acknowledges partial financial support by the Spanish Ministerio de Ciencia e Innovaci\'on Grant No. PID2020–116567 GB-C21 funded
by Grant No. MCIN/AEI/10.13039/501100011033, and
the Project No. PROMETEO/2020/079 (Generalitat
Valenciana).

\end{document}